\documentclass[usenatbib,fleqn,twocolumn]{mn2e}
\usepackage{mathptmx}
\usepackage{ulem}
\usepackage[utf8x]{inputenc}
\usepackage{amsmath,amssymb,amsbsy}
\usepackage[pdftex]{graphicx} 
\usepackage{hyperref}
\usepackage[usenames,dvipsnames]{color}
\usepackage{bm}
\usepackage{breqn}
\usepackage{comment}
\usepackage{natbib}

\definecolor{MyDarkBlue}{rgb}{0,0.1,0.7}
\hypersetup{pdfborder={0 0 0},colorlinks,breaklinks=true,
  urlcolor={blue},citecolor={MyDarkBlue},linkcolor={magenta}}

\title[Radio Emission from Multipolar Magnetars]{Angular Dependence of Coherent Radio Emission from Magnetars with Multipolar Magnetic Fields}

\author[Yamasaki, Ek{\textcommabelow s}i \& G{\"o}{\u{g}}{\"u}{\textcommabelow s}]
{Shotaro Yamasaki\thanks{E-mail: shotaro.s.yamasaki@gmail.com}$^{1,2}$, Kaz{\i}m Yavuz Ek{\textcommabelow s}i$^{3}$, and  Ersin G{\"o}{\u{g}}{\"u}{\textcommabelow s}$^{4}$
\\
$^{1}$Racah Institute of Physics, The Hebrew University of Jerusalem, Jerusalem 91904, Israel\\
$^{2}$Department of Physics, National Chung Hsing University, 145 Xingda Rd., South Dist., Taichung 40227, Taiwan (R.O.C.)\\
$^{3}${\.{I}}stanbul Technical University, Faculty of Science and Letters, Physics Engineering Department, 34956, {\.{I}}stanbul, Turkey\\
$^{4}$Sabanc{\i} University, Faculty of Engineering and Natural Sciences, Tuzla 34956, {\.{I}}stanbul, Turkey
}

\begin{document}

\maketitle

\begin{abstract}
The recent detection of a Fast Radio Burst (FRB) from a Galactic magnetar secured the fact that neutron stars (NSs) with super-strong magnetic fields are capable of producing these extremely bright coherent radio bursts. One of the leading mechanisms to explain the origin of such coherent radio emission is the curvature radiation process within the dipolar magnetic field structure. It has, however, already been demonstrated that magnetars likely have a more complex magnetic field topology. Here we critically investigate curvature radio emission in the presence of inclined dipolar and quadrupolar (``quadrudipolar'') magnetic fields and show that such field structures differ in their angular characteristics from a purely dipolar case. We analytically show that the shape of open field lines can be modified significantly depending on both the ratio of quadrupole to dipole field strength and their inclination angle at the NS surface. 
This creates multiple points along each magnetic field line that coincides with the observer's line of sight, and may explain the complex spectral and temporal structure of the observed FRBs.
We also find that in quadrudipole, the radio beam can take a wider angular range and the beam width can be wider than in pure dipole. This may explain why the pulse width of the transient radio pulsation from magnetars is as large as that of ordinary radio pulsars.
\end{abstract}


\section{Introduction}
\label{sec:intro}

Highly magnetized neutron stars (NSs), in short magnetars \citep{dun92}, have magnetic field strengths above the Schwinger limit $B_{\rm cr}\equiv m_{\rm e}^2c^3/(\hbar e)\sim 4.4\times 10^{13}$ G and slow spin periods of about seconds, and emit recurrent X-ray bursts, which are short in duration but extremely energetic events \citep{kas17,enoto19}. Magnetic field strengths of magnetar sources are generally inferred assuming that the NS loses its energy through the magnetic dipole radiation. Namely, the magnetic field estimates using their spin periods and spin-down rates are the indicators of their dipole field strengths. However, magnetars likely possess multipolar field structures which could play important roles in producing energetic X-ray bursts. For example, the inferred magnetic field strength of SGR 0418+5729 is only $6 \times 10^{12}$~G \citep{rea+13}. However, \citet{guv+11} investigated the persistent X-ray spectrum of this source with a physically motivated spectral model and determined its surface magnetic field averaged over the spin cycle as $10^{14}$~G. They concluded that the stronger fields reside in multipolar components, and likely give rise to energetic bursts. By performing a spin phase-resolved spectral investigation of the same data set \citep{tie+13} found a variable spectral absorption feature in a fraction of the spin cycle. As interpreted as a proton cyclotron feature, the corresponding magnetic field strength was found to be above $2\times 10^{14}$~G, implying the presence of multipolar magnetic topology in a magnetar. Furthermore, although not a magnetar, recent modelling of NICER observations of thermal X-ray pulses from a radio pulsar PSR J0030+0451 has shown a strong preference for high-temperature surface regions created by non-dipolar magnetic field configurations \citep{riley2019,bil+19,kalapotharakos21}. Recent advanced numerical simulations by \citet{igo+21} also indicate to the presence of complicated magnetic field structure of magnetars.

With magnetic field strengths above $B_{\rm cr} $, magnetars have the necessary energy budget and ambient conditions to operate the coherent mechanisms of fast radio bursts (FRBs; \citealt{lor+07,tho+13}), short-duration (typically a few milliseconds) intense (peak flux densities up to about 100~Jy) coherent (brightness temperatures of about 10$^{35}$~K) signals observed in radio bands (GHz), to be observed even from Gpc distances. 
Therefore, the currently popular models to elucidate FRBs generally invoke NSs with magnetar type field strengths \citep[see e.g., ][]{beloborodov20}. 
A Galactic magnetar, SGR J1935+2154 has entered into a burst active episode on 2020 April 27, emitting hundreds of energetic X-ray bursts in a day \citep{pal20}. During this intense burst activity phase, a bright radio burst from this magnetar was detected independently with two radio facilities: CHIME \citep{chimegalacticfrb} and STARE2 \citep{boc+20}. 
The detection of FRBs \citep{stare2_galacticfrb,chimegalacticfrb} associated with its X-ray bursts \citep{integral_paper,hxmt_paper,konus_paper,agile_paper} from a Galactic magnetar SGR J1935+2154 establishes that at least some FRBs originate from magnetars. 

\begin{figure*}
\includegraphics[width=0.45\linewidth]{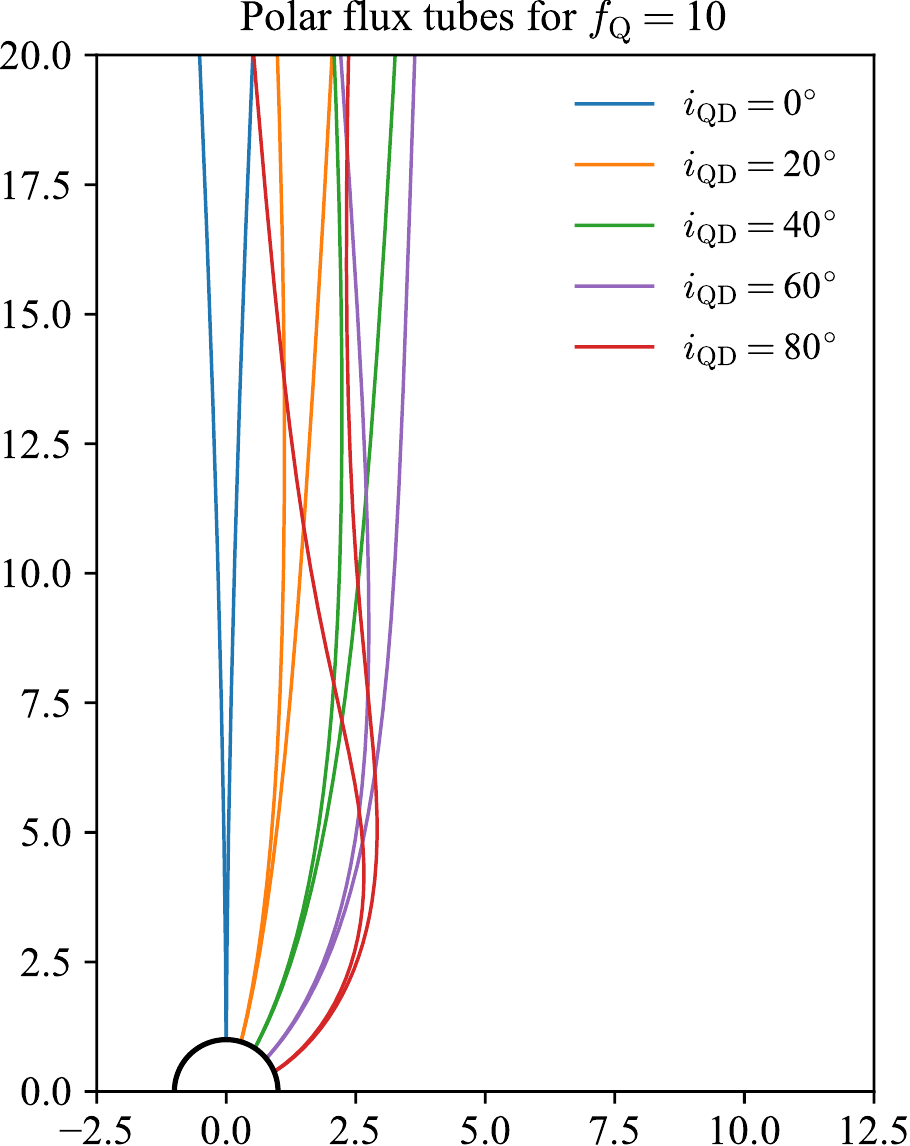}
\hspace{0.5cm}
\includegraphics[width=0.45\linewidth]{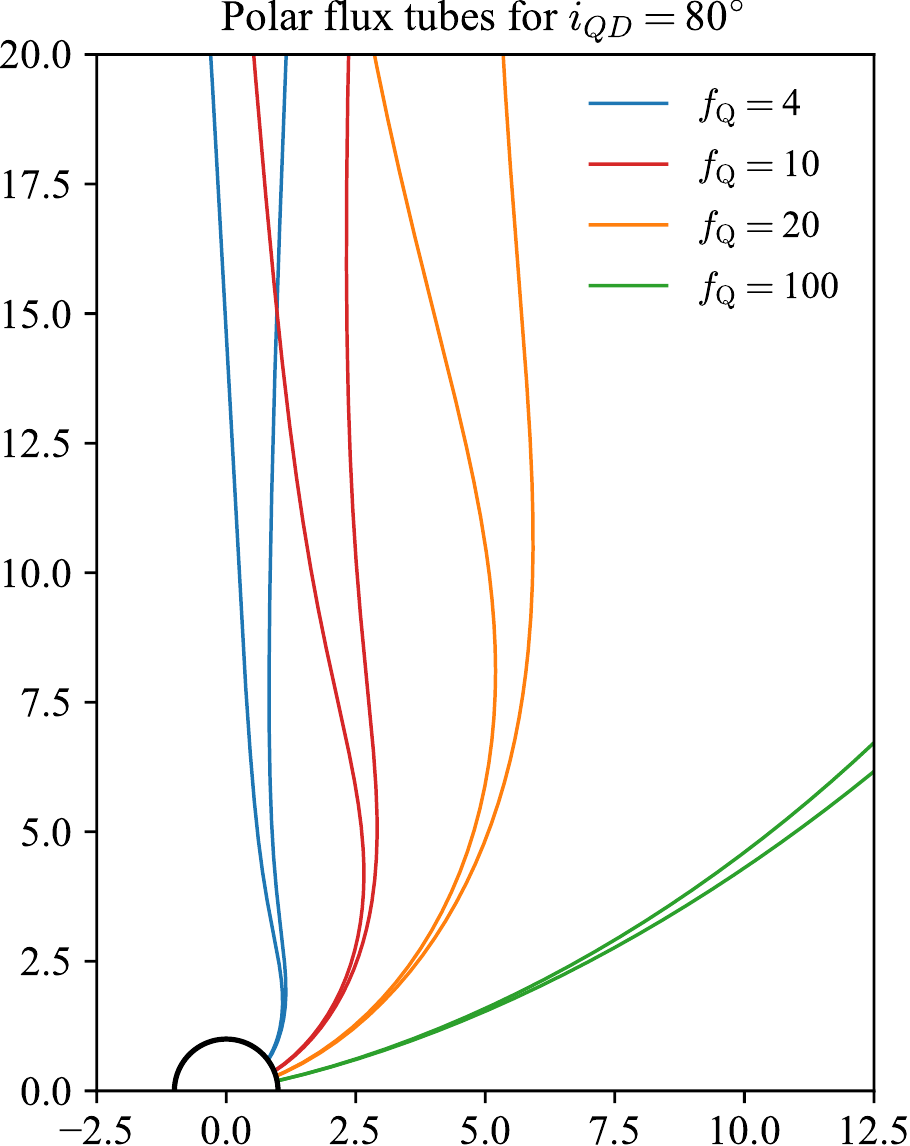}
\caption{Polar flux tubes for different inclinations with $f_{\rm Q}\equiv B_Q/B_D=10$ (left) and $i_{QD}=80^{\circ}$ (right). The assumed parameter for the NS is $x_{\rm LC}\equiv R_{\rm LC}/R_\ast=2\times10^4$ (i.e., $P\sim 4.2$~s), where $R_\ast=10^6$~cm. The expected geometry significantly deviates except for the aligned quadrupole and dipole case ($i_{QD}=0^{\circ}$) in the left panel, which is almost identical to a pure dipole one. The case of $f_{\rm Q}=10$ and $i_{QD}=80^\circ$ (reddish curves) is identical between left and right panels for reference.}
\label{fig:oblique}
\end{figure*}

Recent observations of repeating source FRB 20121102A revealed that the distribution of burst waiting times is bi-modal with peaks at $\sim3$ ms and $\sim70$ s \citep{li21,hewitt21}. The existence of a burst population with such a short waiting time may favour the magnetospheric origin of bursts \citep{li21}. Among these, there are models for FRBs in which some sort of magnetic disturbance near the NS, such as stellar quake driven Alfv\'en waves \citep{katz16,kum+17,ghi17,yang18,lu20}, trigger coherent curvature radiation along open magnetic field lines in the \textit{dipolar} magnetosphere. 
Meanwhile, as evidenced by observations, the NS population generally could possess multipolar field components, which can affect emission processes occurring near the star and may result in the modified angular characteristics of coherent radio emission from magnetars.
For instance, \citet{yang21} recently show schematically that the diversity in the burst occurrence rate of FRBs might have to do with the potential existence of multipolar field geometry near the star in the conjuncture of the Galactic FRB-like bursts from SGR 1935+2154 (see also \citealt{wang21}). Since the presence of such multipolar fields might significantly affect the curvature of the magnetospheric field lines in the emission zone, it must be taken into account when modelling and interpreting the observed emission. 

In this work, we show with simple but realistic multipolar magnetic field configurations how angular characteristics for observing transient coherent radio emission (including FRBs and transient pulsed radio emission) from magnetars are modified assuming that they could arise from the curvature emission along the open multipolar magnetic fields. 
In particular, we consider the ``quadrudipolar'' field configuration in which the quadrupolar component is superimposed onto the dipolar component with an arbitrary field strength ratio and inclination angle between two components (\citealt{BA1982}, hereafter BA82). 
In general, a quadrudipolar field configuration yields two open field line regions at each pole. The open field line region in the northern hemisphere (or ``polar cap'') has a nearly circular shape whereas that in the southern hemisphere exhibits a thin annular shape when dipolar and quadrupolar moments are aligned, and it could be much more complicated in cases they are misaligned (BA82, see also \citealt{gralla17} for detailed numerical simulation of an aligned quadrupole). 
As a first step toward understanding the general field configuration, we consider the radio emission from the Northern\footnote{Northern hemisphere is defined by the region where the inner product of the dipole moment and the normal vector at the stellar surface is positive.} polar cap, which has a larger area at the stellar surface than the Southern one and presumably a higher ability to generate observable radio emission \citep{gralla17,lockhart19}. Although our ultimate goal is to study transient radio emissions from magnetars using the force-free field, which could approximate the magnetic field of the plasma-filled magnetosphere, in this paper we concentrate on the qualitative investigation of angular characteristics of radio beams using vacuum
magnetic field only.

This paper is organized as follows. In \S~\ref{sec:model} we present the quadrudipolar field configurations that we use. In \S~\ref{sec:implication} we discuss the implications of quadrudipole for FRBs, and in turn for transient radio pulses from magnetars. Finally, we conclude by summarizing our findings and prospects in \S~\ref{sec:discuss}.

\section{Modification of magnetic geometry by quadrupole fields}
\label{sec:model}

We consider the emission geometry from a NS by solving the viewing geometry in an inclined dipolar and quadrupolar magnetic field following BA82. 
We will use the vacuum field as an approximation to the magnetic field of the plasma-filled magnetosphere, where the magnetospheric structure is stationary in the corotating frame.
Important model parameters are the surface field strength ratio between quadrupole and dipole {$f_{\rm Q}\equiv B_{\rm Q}/B_{\rm D}$} and the inclination angle between dipole and quadrupole moments $i_{\rm QD}$. 

A generic limit on the quadrupolar field strength comes from, of course, the condition $GM_{\ast}^2/R_{\ast}^4>B_{\rm Q}^2/(8\pi)$, where $R_{\ast}$ and $M_{\ast}$ are the stellar radius and mass, respectively. For a typical magnetar, this gives $f_{\rm Q}<3.6\times 10^3\  (M_{\ast}/1.4 M_{\odot})\,R_{\ast,6}^{-2}\,B_{\rm D,15}^{-1}$, where $B_{\rm D,15}\equiv B_{\rm D}/(10^{15}\,{\rm G})$ and $R_{\ast,6}\equiv R_{\ast}/(10^{6}\,{\rm cm})$.
The quadrupolar field strength may also be constrained by a spin-down theory. In principle, a strong quadrupolar field component could affect the evolution of NSs in terms of the spin-down. The ratio between the pure dipole and quadrupole losses is estimated as 
\begin{equation}
    \frac{L_{\rm Q}}{L_{\rm D}}=\frac{32}{9}f_{\rm Q}^2\ x_{\rm LC}^{-2}\sim 1.6\times10^{-7}\, R_{\ast,6}^{2}\ f_{\rm Q}^2\left(\frac{ P}{1\ {\rm s}}\right)^{-2}
\end{equation}
\citep{petri15}, where $x_{\rm LC}\equiv R_{\rm LC}/R_{\ast}$ with $R_{\rm LC}\equiv c P/(2\pi)$ being the light cylinder radius, and $P$ is the spin period of NSs. This suggests that the effect on the spin-down rate of a slowly rotating NS (at $P\sim1$ s) due to the existence of quadrupole component is relevant only when the quadrupole to dipole field strength ratio is extremely high $f_{\rm Q}\gtrsim 10^3$.  Consequently, we consider a typical range of $f_{\rm Q}\lesssim10^3$ throughout\footnote{In contrast, the assumption that the dipole is dominant over the quadrupole (i.e., $L_{\rm Q}/L_{\rm D}\lesssim1$) for typical radio pulsars that are rotating fast, although not trivial, severely constrains $f_{\rm Q}$. For instance, this yields $f_{\rm Q}\lesssim 12$ for a radio pulsar PSR J0030+0451 with spin period of $4.9$ ms.}. Regarding the inclination angle, we mainly focus on oblique cases with $i_{\rm QD}>0$ because it is expected to enhance the deviation from the pure dipole and in general, there is little reason to assume an aligned case with $i_{\rm QD}=0$.

\begin{figure*}
\includegraphics[width=0.48\linewidth]{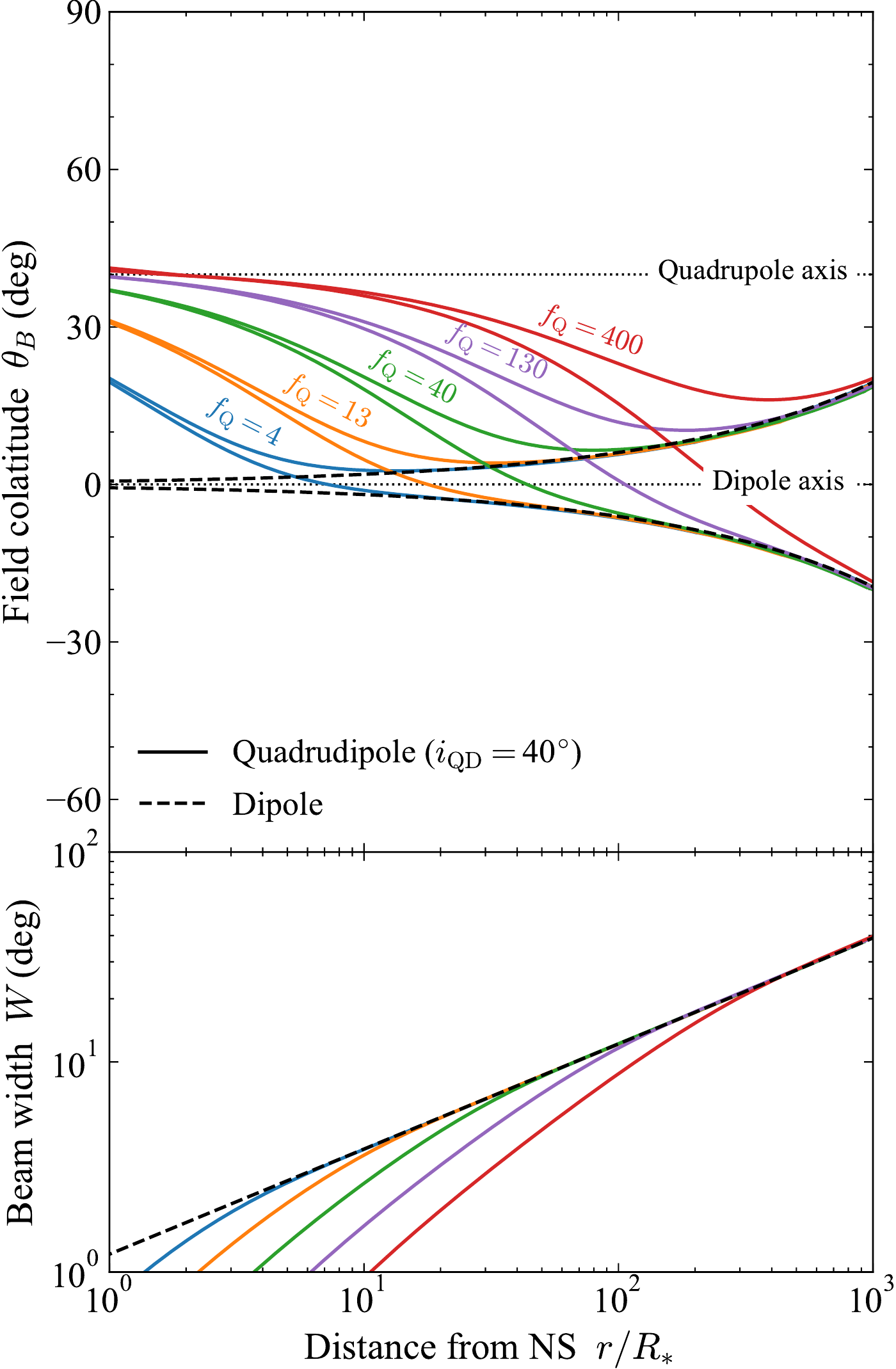}
\hspace{0.5cm}
\includegraphics[width=0.48\linewidth]{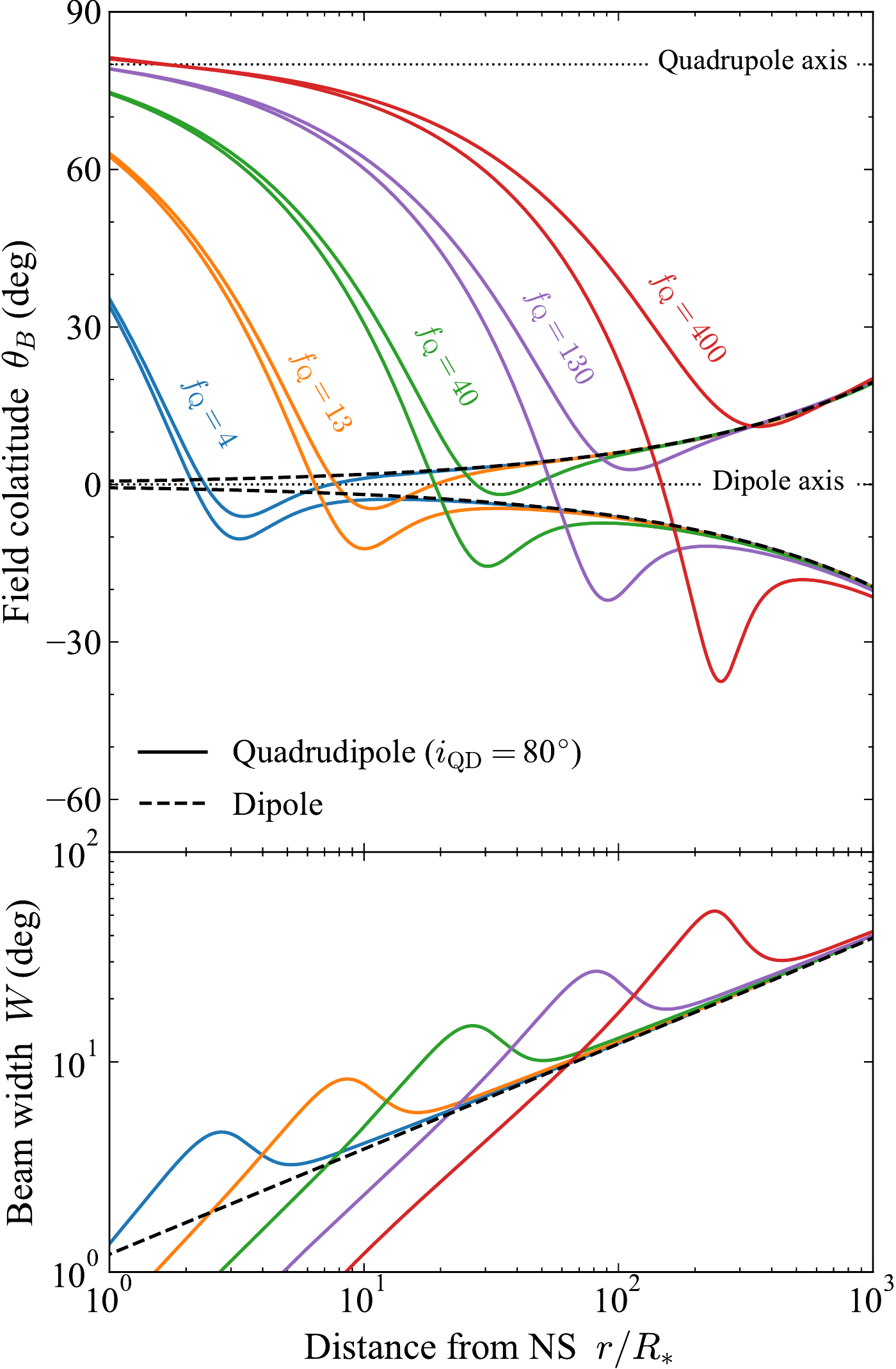}
\caption{{\it Top:} Colatitude of the last open field lines as a function of distance from the star for a pure dipolar geometry (dashed lines) and for quadrudipolar geometries with $i_{\rm QD}=40^{\circ}$ (left) and $i_{\rm QD}=80^{\circ}$ (right) when varying $f_{\rm Q}\equiv B_{\rm Q}/B_{\rm D}$ from $4$ to $400$ (solid lines). In the extreme limit of $f_{\rm Q}\rightarrow0$, the quadrudipole asymptotically approaches the pure dipole. {\it Bottom:} the beam width $W$, defined by the angular extent of last open field lines, as a function of radius. The assumed parameters for the NS here are the same as in Figure \ref{fig:oblique}. 
}
\label{fig:theta_B}
\end{figure*}

\subsection{Field Geometry}

The components of the total magnetic field are generally written as
\begin{equation}
\label{eq:B-field superposed}
\begin{aligned}
B_r &= B_{\rm D}\left[ \frac{\cos{\theta_{\rm D}}}{x^3}+ \frac{f_{\rm Q}}{2} \frac{(3\cos^2{\theta_{\rm Q}}-1)}{x^4} \right] \\
B_{\theta} &= B_{\rm D}\left[ \frac{\sin{\theta_{\rm D}}}{2\,x^3}+ f_{\rm Q} \frac{\cos{\theta_{\rm Q}}\sin{\theta_{\rm Q}}}{x^4} \right]
\end{aligned}
\end{equation}
where $x$ is the radius normalized by the stellar radius.
Hereafter, we express the coordinate in terms of the dipole, i.e., $\theta_{\rm D}=\theta$ and $\theta_{\rm Q}=\theta-i_{\rm QD}$. 
In order to find the polar flux tube, the differential equation
\begin{equation}
\label{eq:B-flux conservation}
\frac{{\rm d} r}{B_r} = \frac{r\, {\rm d} \theta}{B_{\theta}}
\end{equation}
can be solved by integrating the field line equation \eqref{eq:B-field superposed} from the equatorial point at the light cylinder radius $r=R_{\rm LC}$, $\theta=\pm \pi/2$ to the surface for the two last open field lines. A fully analytic solution for the field lines are given in BA82 for the aligned case ($i_{\rm QD}=0$). An analytic solution of \autoref{eq:B-flux conservation} which is valid at $x\sin{\theta}\gg f_{\rm Q}$ is also given for the general obliquity ($i_{\rm QD}\neq 0$):
\begin{dmath}
    x=\sin^2{\theta} \, \left \{ x_{\rm eq}
    +\xi f_{\rm Q}\, S_1+f_{\rm Q}\left[\left(\frac{5C_1-1}{4}\right)\frac{\cos{\theta}}{\sin^2{\theta}}+\left(\frac{1-C_1}{4}\right)\ln\left|\frac{1-\cos{\theta}}{\sin{\theta}}\right|+S_1\left(\frac{1}{3\sin^2{\theta}}-\frac{5}{\sin{\theta}}\right)\right]\right \},
\end{dmath}
where $C_1=\cos{(2i_{\rm QD})}$ and $S_1=\sin{(2i_{\rm QD})}$, and $x_{\rm eq}$ is the field line constant which satisfies the boundary condition of $x=x_{\rm eq}$ at $\theta=\pm \pi/2$. Here, $\xi=14/3$ for $\theta=\pi/2$ and $\xi=-16/3$ for $\theta=-\pi/2$\footnote{We correct the minor mistake ($\xi=-14/3$ for $\theta=-\pi/2$) in BA82, although this barely affects the resulting field geometry.}. 
Near the star ($x\sin{\theta}\ll f_{\rm Q}$), this analytic solution is inaccurate and should be continued to the surface of the NS by numerical integration. \autoref{fig:oblique} shows the two last open field lines $x_{\rm eq}=x_{\rm LC}$ for different $i_{\rm QD}$ and $f_{\rm Q}$. It is clear that the open field line geometry near the star ($x\sin{\theta}<f_{\rm Q}$) can significantly deviate from a pure dipolar one for high inclination ($i_{\rm QD}$) and/or high quadrupole to dipole field strength ratio at the surface ($f_{\rm Q}$). Meanwhile, at larger radii ($x\sin{\theta}\gg f_{\rm Q}$) any field line approaches to the pure dipolar field geometry. 

\subsection{Beam Characteristics}
\label{ssec:beam}
Here we discuss the angular characteristics of radio emission. \autoref{fig:theta_B} shows the polar angle of last open field lines, $\theta_{\rm B}$, and the angular width between them, $W$, as a function of radius $r$ from the NS for different field geometries. There are two effects of including the quadrupolar component: significant deviations of (i) the polar angle of the emission and (ii) the width of the flux tube (i.e., the radio beam) from those in a pure dipolar field geometry.

First, the polar angle of quadrudipolar field has a stronger dependence on $r$ than that a dipole. 
For a pure dipolar case, the polar angle monotonically increases with $r$, and thus there is always a single solution of $\theta_{\rm B}(r)=\theta_{\rm obs}$ for $r$ where $\theta_{\rm obs}$ is the observer viewing angle measured from the north pole. For a quadrudipolar case, on the other hand, each field line has more than one solution at which the local field line is directed towards the observer, satisfying $\theta_B(r_i)=\theta_{\rm obs}$ (see top panels of \autoref{fig:theta_B}). Specifically, a field line passing through the equatorial point at $\theta=\pi/2$ ($\theta=-\pi/2$) could have at most two (three) solutions. 
Since a field line at locations corresponding to such solutions have different curvatures, this may lead to complex temporal and spectral radio pulse profiles.

Second, multipolar fields generally have much narrower opening angles at the surface than dipole fields (BA82). However, this does not necessarily mean that chance probability of seeing radio emission from the multipolar fields is lower than that from pure dipole. In the case of quadrudipolar fields, the angular extent between the polar flux tubes near the point where a sign of curvature changes $r/R_\ast\sim f_{\rm Q}$ could be even larger than the dipolar one for high inclinations $i_{\rm QD}\gtrsim 50^\circ$  (see the bottom right panel of \autoref{fig:theta_B}). Moreover, as a polar angle of quadrupolar field lines is a strong function of radius it could have a larger coverage of the sky below given radius than a purely dipolar one.

Special relativistic effects, such as the aberration of photon emission direction and photon travel time delay,
could be in principle important near the light cylinder, where the structure of last open field lines is modified \citep[e.g.,][]{yadigaroglu97}. However, it does not significantly change the structure of inner magnetosphere (at radii below $f_Q \,R_\ast\sim R_{\rm LC}/100$ for $P=1$ s magnetars), where the quadrupole plays important roles.

\section{Implications for transient radio emissions from magnetars}
\label{sec:implication}
Now we turn to discuss the implications of multipolar fields for the coherent radio emission from magnetars, including FRBs and transient radio pulsations from galactic magnetars \citep{cam+06,cam+07}, in the framework of curvature emission from multipolar field geometries\footnote{Although the evolution of the magnetic fields in NSs largely remains unclear, except for multipoles of very high order $n$, it is not that different from that of a pure dipole when assuming dissipation mechanisms (the dissipation timescale differs only about $n$ times), such as the Ohmic dissipation (e.g., \citealt{kro91,arons93,mitra99,igoshev16}). Namely, the characteristic timescale for the emergence or disappearance of quadrupolar fields is typically longer than the activation timescale of magnetar (from months to years) during which (at least the Galactic) FRBs and coherent radio pulsation preferentially occurs. }. 
The curvature emission frequency is estimated as $\nu\propto\Gamma^3/R_{\rm c}$, where $\Gamma$ is the Lorentz factor of electrons in bunch and $R_{\rm c}$ is the curvature radius. Although the demand that the emitted frequency is in the radio band ($\sim$GHz) constrain $\Gamma$ and $R_{\rm c}$, a full determination of them requires self-consistent modelling of particle generation and acceleration (e.g., \citealt{kum+17,ghi17,yang18,kumar20,lu20}), which remains unclear (see, e.g., \citealt{lyubarski21} for criticisms). 
Instead, we consider only the geometrical effects of magnetic fields on the emission direction assuming that coherent radio emission is successfully generated at some radius.

The observed emission is not isotropic, but it is beamed within some fraction of solid angle $4\pi$. 
In case of quadrudipolar field structure, the maximal timescale for the observer's line of sight (LOS) to sweep the beam is typically
$T_{\rm beam}=PW/(2\pi)\sim110 {\ \rm ms}\ (P/4{\,\rm s})(W/{\cal O}(10{\,\rm deg}))$,
where $W$, of course, depends on $r$, $f_{\rm Q}$, and $i_{\rm QD}$ (see the bottom panels of \autoref{fig:theta_B}). 

The trigger timescale is also important in determining the observed duration of coherent emission $t_{\rm obs}$.
FRBs are believed to be generated by a series of instantaneous and stochastic triggers such as occasional starquakes. Namely, FRBs would be observed only when the magnetic field lines (or the beam), along which instantaneous generation and acceleration of charged particles take place, cross the observer's LOS. Meanwhile, the radio pulsations from magnetars, although highly variable, could be almost steadily triggered during the active period of magnetars.  
Therefore, the observed duration of FRBs is controlled by intrinsic trigger timescale $t_{\rm obs}=t_{\rm int}/(1+z)$ (with $z$ being the source redshift), whereas the observed pulse width of radio pulsation from magnetars is determined by the geometric beam size $t_{\rm obs}=T_{\rm beam}$ when neglecting the effects of the potential substructure inside the beam and the impact parameter of the LOS with respect to the beam center.

\subsection{Fast Radio Bursts}

Intriguingly, a viewing angle configuration with multiple emission points may lead to an FRB with complex temporal structure. The observed temporal separation between 
different emission components would be estimated by considering the differences both in emission times and light travel times \citep{wan+19drift,wan+20,lyut20-map}\footnote{Note that here a time delay arises from two emissions along the {\it same} quadrupolar field line whereas in \citet{wan+19drift,wan+20,lyut20-map} it arises from two emissions along {\it different} dipolar field lines. The arrival time delay in the latter is even smaller than in the former and thus negligible.}:
\begin{equation}
    \Delta t
    \approx \frac{\Delta r}{c} \,\left(1-\cos\Delta \theta+\frac{1}{2\Gamma^2}\right) \sim 0.5{\ \rm ms}\ \left(\frac{\Delta r}{10^2\,R_\ast}\right)\left(\frac{\Delta \theta}{30^\circ}\right)^2\ ,
\end{equation}
where $\Delta r$ is the distance between two emission points, $\Delta\theta$ is the angle that the line passing through the two emission points makes with respect to the LOS. Both $\Delta r$ and $\Delta\theta$ increase as $f_{\rm Q}$ increases, but they are not much larger than quoted values. The relativistic effects of motion along the LOS is neglected in the second equality assuming $\Gamma\gg1$. Therefore, multiple solutions most likely manifest themselves as a burst consisting of either separable or inseparable multiple-components. 
These might explain observations of some repeating FRBs with complex time-frequency structures \citep{chime19Natur,chime19ApJ,josephy19,hes+19} and thus-far non-repaeting FRBs with double-peaked features \citep{champion16,cho20,day20}.

In the presence of quadrudipolar field, a polar angle of field line is a strong function of $r$. This means that even a small variation in emission altitude could lead to the large variation in the observed emission angle. This would naturally give rise to the highly sporadic burst rate of repeating FRBs.

In most FRBs, the time-frequency structure exhibits a downward drifting pattern, i.e., the sub-pulses arriving later have lower frequencies with some exceptions (e.g., \citealt{pleunis21}).
Theoretical models for frequency drift patterns are already complex even in a pure dipolar case \citep[e.g.,][]{wan+19drift,wan+20,lyut20-map}. Including the quadrupole gives further freedom in generating more complex frequency drift patterns, which may explain the observed diversity of frequency drift. Also, the inclusion of multipolar field should alter the radius-to-frequency mapping that assumes a pure dipole \citep[e.g.,][]{lyut20-map,wang21} due to the non-monotonic behavior of the field line curvature.

\subsection{Transient Radio Pulsation from Magnetars}

To date, coherent radio pulsations have been detected from six\footnote{Note that this does not include a NS PSR J1119-6127 which exhibited magnetar-like bursting/flaring activities but has relatively weak dipolar magnetic field strength of $\sim10^{13}$ G for classic magnetars.} (including SGR 1935+2154, the source of the Galactic FRB) of the 30 currently known magnetars \citep{olausen14}\footnote{\url{http://www.physics.mcgill.ca/~pulsar/magnetar/main.html}}. All of the six sources are transients, and the radio emission occurred preferentially when they exhibited bursting/flaring activities in X-rays. Although the origins of transient radio pulses of magnetars remain unclear, it has also been proposed that they may be generated in the closed magnetic field region of a dipolar configuration \citep{wad19,wang19_magnetar}.
Such models invoking acceleration along closed field lines are partly motivated by the fact that the angular size of open field lines in a pure dipole tends to be extremely small for magnetars rotating slowly (beam size of pure dipole scales with $\propto P^{-1/2}$) while that of closed field lines could be arbitrarily large. Another motivation comes from the fact that the relatively hard spectra of magnetar radio pulsation extending up to a few tens of GHz is better explained by the emission from closed field lines which are more curved than open field lines.

Alternatively, our results imply that magnetar radio pulsations may arise from the open field region modified by multipolar components, as suggested by early polarization observations \citep{kramer07}, for the following reasons.
First, as shown in \S \ref{ssec:beam}, open field lines of a quadrudipole with high obliquity could have a wider beam size at a given radius and cover a larger fraction of the sky below a given radius than those of a pure dipole do (the latter point is critical since it significantly increases the effective beam size). 
Indeed, the duty cycle of radio pulsation from a magnetar XTE J1810-197 with spin period of $5.5$ s \citep{ibrahim04} is about $5$\% (translated in beam size of $W\gtrsim18^{\circ}$) at $2$–$8$ GHz range \citep{camilo16,eie21} which is comparable to the duty cycle of ordinary millisecond pulsars \citep{maciesiak11}. In this respect, the generation of radio emission in the open field region with multi-polar configuration could also be a viable option to have a wide enough beam width. Secondly, the open field lines modified by multi-polar field component naturally give rise to the hard spectra of magnetar radio pulsations since its curvature could be effectively smaller than a pure dipole if $\Gamma$ is constant along the field lines.
Finally, the magnetar radio pulsation is known to be highly variable in time, which may imply that the physical condition $\Gamma$ and $R_{\rm c}$ could change dramatically from one pulse to another. From the same reason, we predict that the highly complicated beam structure would result in the significant dispersion in the peak position of the radio pulse profile.

\section{Concluding Remarks}
\label{sec:discuss}

In this work, the magnetic field geometry of an inclined dipolar and quadrupolar magnetic field is considered based on analytic model by BA82 in conjunctures with coherent radio emission from magnetars. While we do not attempt to explain the phenomena at the emission level, we demonstrate that a consideration of multipolar component could give considerable freedom in interpreting/modeling observations of a family of coherent emissions from magnetars, FRBs and transient radio pulsations.

We analytically demonstrate that the curvature of the open field lines can vary significantly depending on both the ratio of quadrupole to dipole field strength and their inclination angle at the NS surface. This means that there 
are multiple points along each magnetic field line that coincide with the observer's line of sight, and may explain the complex spectral and temporal structure of the observed FRBs. This also implies that even a small variation in emission altitude could result in a large variation in the observed emission angle, leading to a highly sporadic burst rate of repeating FRBs.
It is also found that in quadrudipole, the radio beam can take a wider angular range and the beam width can be wider than in pure dipole. This may explain why the pulse width of the transient radio pulsation from magnetars is as large as that of ordinary radio pulsars.

The fact that the magnetar SGR~1935+2154, which is the source of Galactic FRB 200428, has now turned into radio pulsar \citep{fast_atel}, makes it interesting to continue monitoring with high sensitivity radio telescopes. Although testing whether non-dipolar field components are involved only from radio observations may be difficult, it is important for analysis of radio emissions from magnetars to keep in mind that the magnetic field topology may well deviate from the simple dipole model.


\section*{Acknowledgements}
We thank Shota Kisaka for useful comments and discussion. We also thank the anonymous referee for their careful reading of the manuscript and suggestions. SY was supported by the advanced ERC grant TReX. KYE acknowledges support from T{\"U}B{\. I}TAK with grant number 118F028.

\section*{Data Availability}
This is a theoretical paper that does not involve any new data. The model data presented in this article are all reproducible.

\footnotesize{
\bibliographystyle{mn2e}
\input{multipole.bbl}
}

\end{document}